\documentclass[12pt]{article}
\usepackage{epsfig,graphicx}

\begin{document}
\title{Persistence in Practice}
\author{J.M.J. van Leeuwen, \\
Instituut-Lorentz, University of Leiden, P.O.Box 9506, \\
2300 RA Leiden, the Netherlands\\*[4mm]
V.W.A. de Villeneuve and H.N.W. Lekkerkerker, \\
Van 't Hoff Laboratory for Physical and Colloid Chemistry, \\
University of Utrecht, Padualaan 8, \\
3584 CH Utrecht,  The Netherlands}
\maketitle

\begin{abstract}
We present a scheme to accurately calculate the persistence probabilities
on sequences of $n$  heights above a level $h$ from the measured $n+2$ points
of the height-height correlation function of a fluctuating interface. The calculated
persistence probabilities compare very well  with the measured persistence probabilities
of a fluctuating phase-separated colloidal interface for the whole experimental range.
\end{abstract}
\maketitle
\noindent

\section{Introduction}

Persistence concerns the question how long a fluctuating variable stays above a certain level. 
It is a recurrent theme in statistical physics. It started as the problem of level crossing in probability
theory \cite{Slepian,Kac,Blake}. If the problem is posed in the context of Gaussian random processes,
persistence is completely determined by the correlation function of the stochastic variable. 
For these processes the essence of the problem is to calculate, from the correlation function, the
probability that the fluctuation variable stays above a certain level during the time $t$.
This turned out to be a classic unsolved problem in probability theory \cite{Kac,Sire,Krug}.
Around the turn of the century the investigations reached a peak, with applications 
in physics, ranging from properties of the diffusion equation \cite{Cornell,Newman,Hilhorst},
survival of spin states \cite{Derrida,Hakim,Zeitak,Sire,Majumdar}, fluctuating steps 
\cite{Dougherty} and interfaces \cite{Krug,Kallabis}, 
to persistence in order parameters \cite{Oerding}  and non-Gaussian processes \cite{Deloub}.
A recent review on equilibrium step fluctuations, which poses the problem of persistence in
a wide context,  is given by Constantin et al. \cite{Constantin}.

The investigations focussed on the calculation the asymptotic behavior. The associated
persistence exponent sometimes characterizes a new partition of physical systems into 
universality classes.  For a stationary Gaussian process the persistence probability for a time 
$t$ decays exponentially with $t$. For these processes
the persistence probability can be written as the ratio of two path integrals. The numerator 
involves the sum over paths obeying the condition of persistence and the normalizing denominator 
over paths without the condition. For a Markov process the correlation function is an exponential and 
both path integrals can be evaluated.  But as yet there is no general scheme for an arbitrary correlation
function to calculate the persistence exponent. 

Most theoretical investigations treat the process as a continuous process in time, which it certainly is.
However measurements of a stochastic variable are necessarily discrete in time. 
One would think that, with a sufficiently high experimental sampling rate, the limit of a continuous 
process would be seen experimentally, such that the sampling rate would become irrelevant. 
This is not obvious though, in particular for "non-smooth" processes \cite{Sire}. 
A "smooth" process has a correlation function deviating from  its initial value in a quadratic way. 
The influence of the discreteness of the sampling on the persistence exponent
has been investigated by Majumdar et al. \cite{Majumdar2} for stationary Gaussian Markov 
processes and by Ehrhardt et al. \cite{Ehrhardt} for a number of non-Markovian smooth processes.
In general the exponent derived from discrete sampling is lower than the continuous exponent,
because double crossing of the level in between two discrete sampling points are missed. 
These authors consider the calculation of the persistence of a discrete sequence 
more difficult than that of the underlying continuous process, as the approximation schemes derived for
continuous processes are not applicable to discrete series of data. 
For "non-smooth" processes another difficulty arises in the calculation of the persistence probability, 
because rapid fluctuations give a diverging probability on a short time scale, such that the mean
persistence time vanishes \cite{Krug,Volkert}. 

Recently we encountered the persistence problem in an experimental study of a fluctuating 
interface between phase-separated colloid-polymer mixtures \cite{Volkert}. 
This is an example of a stationary Gaussian random process, since the fluctuations are small in 
amplitude and their energy is given by a quadratic hamiltonian in the stochastic variables. 
On the time scale of the measurements the process is non-Markovian and non-smooth. 
We have collected data on the correlation function and on the joint probabilities for e.g.
finding $n$ successive height values above the level $h$. Thus here the practical persistence
problem presents itself in a discretized form and the question arises whether
one can calculate, from the values of the correlation function, directly the persistence
probabilities of the same set of points in time. In an earlier publication \cite{Hans}
we demonstrated how to do this for short series. In this paper we extend the calculation
to the whole set of experimental points.
We focus here on the longer time sequences which allow for 
more detail than the similar spatial sequences. We find the values of the correlation
function at equidistant time intervals and also the persistence probability 
for the same series of time steps. 

Thus experimental data were obtained in large numbers and high accuracy for an interesting system,
which enable to further test the theories on the persistence problem.
As mentioned in \cite{Ehrhardt}, the continuum limit usually leads to a simplification of the 
calculation, but we find in our case the discreteness of the experimental data rather a blessing
in disguise. Our aim is to show how the persistence probabilities for $n$ events can be directly 
calculated from the measured $n+2$ points of the correlation function for the range of 
measured points, which are neither on a short time scale nor in a fully asymptotic regime. 
We derive two sum rules for the discrete series, which play a vital role in assessing the accuracy 
of the computational scheme.

\section{The experimental system}

Traditionally experimental studies of interfaces are carried out by means of
light and X-ray scattering. The field obtained another dimension by experiments
of Aarts et al. \cite{Aarts,Tanaka}, in which they obtained microscopic images of fluctuating
interfaces of phase separated colloid-polymer mixtures using confocal microscopy.
Although scattering on interfaces is most valuable,
it always yields {\it global} information on the fluctuations, while 
inspection by microscopy gives {\it local} information. However, the wave lengths and 
heights involved in the capillary waves of molecular fluids are way out of the reach of
detection by microscopic methods. For colloidal interfaces the characteristic
length and time scale of the fluctuations can become accessible by confocal microscopy,
by lowering the surface tension to the nN/m range.  
 
Here, confocal microscopy measurements were performed on phase separated
colloid-polymer mixtures. The colloids are 69 nm radius fluorescently labeled
polymethylmetacrylate particles, suspended in cis/trans decalin, with
polystyrene (estimated radius of gyration = 42 nm) added as depletant polymer.
Due to a depletion induced attraction these mixtures phase separate at
sufficiently high colloid and polymer volume fractions and a proper colloid to
polymer aspect ratio, into a colloid-rich/polymer-poor (colloidal liquid) and a
colloid-poor/polymer-rich (colloidal gas) phase \cite{Lekkerkerker}. Here the polymer 
concentration acts as an inverse temperature and upon dilution the binodal is approached.
 
In confocal microscopy a monochromatic laser beam is used to excite dye molecules 
attached to (in this case) the colloid. Through a dichroic mirror the outgoing light is 
separated from the incoming light. A two-dimensional confocal slice is then obtained 
through a pinhole, from single-wavelength fluorescent light emitted from the sample.
For our experimental system, the confocal slices are only ~3 colloidal diameter thick, 
the density profile between the two phases is observed as a function of fluorescent intensity.

A very precise location of the interface can be obtained by fitting the 
intensity with a van der Waals profile: $I(z,x) = a + b \tanh ([z-h(x)]/c)$, 
where $z$ is the direction perpendicular to the interface and $x$ a 
coordinate along the interface. In the upper phase the density approaches 
a value corresponding to $a+b$ and in the lower phase to $a-b$, while $c$ 
measures the intrinsic width of the interface.
Thus at every snapshot a function $h(x)$ follows and the sequence of snapshots
gives the function $h(x,t)$. It is a practical separation of the particle 
motions, which lead at short scales to the {\it intrinsic interface} and the 
particle motions ({\it capillary waves}) which drive the long wavelengths. 
This opens up the possibility to follow in detail the motion of the height of the interface and
to make a statistical analysis of its temporal behavior. Of course the method
has its inherent restrictions. Just as in ordinary movie recording, snapshots
have to be taken at a finite time intervals. For colloidal interfaces this interval
can be made much smaller than the intrinsic time scale of the motions.

With a Nikon E400 microscope equipped with a Nikon C1 confocal scanhead, series 
of 5000 snapshots of the interface were recorded at constant intervals 
$\Delta t$ of 0.45 s and 0.50 s of two statepoints to be denoted as II and IV. 
The latter is closer to the binodal.  The pixels are separated by a distance 
$\Delta x = 156nm$ and a single scan takes approximately 0.25 s to complete. 
With 640 heights per snapshot, we obtain in total $640 \times 5000$
data points, which enable us to measure persistence probabilities as low as $10^{-6}$.

\section{The correlation function}
We write the normalized correlation function as
\begin{equation} \label{a1}
g(t/t_c) =\langle h(0,t) h(0,0) \rangle / \langle h^2 \rangle,
\end{equation}
where $\langle h^2 \rangle$ is the equal time correlation function. 
$t_c$ is the characteristic time of the process which we have inserted in the argument 
of $g$ to make it dimensionless. It is given by the expression
\begin{equation} \label{a2}
t_c = {(\eta+\eta') \over \sqrt{g \gamma\Delta \rho}},
\end{equation} 
where the $\eta$'s are the viscosities of the two coexisting phases, $\Delta \rho$
the density difference and $\gamma$ the surface tension. 
From the definition follows that $g(0)=1$.  Heights are scaled with 
$\langle h^2 \rangle^{1/2}$ and times with 
$t_c$, such we can work exclusively with dimensionless quantities.
Experimentally one has a discrete sampling of $g(t/t_c)$
with $\Delta t$ as smallest interval. Thus one finds a sequence
$g_n = g(n \delta )$ with $\delta=\Delta t / t_c$.

\begin{figure}[h]
      \epsfxsize=\linewidth \epsffile{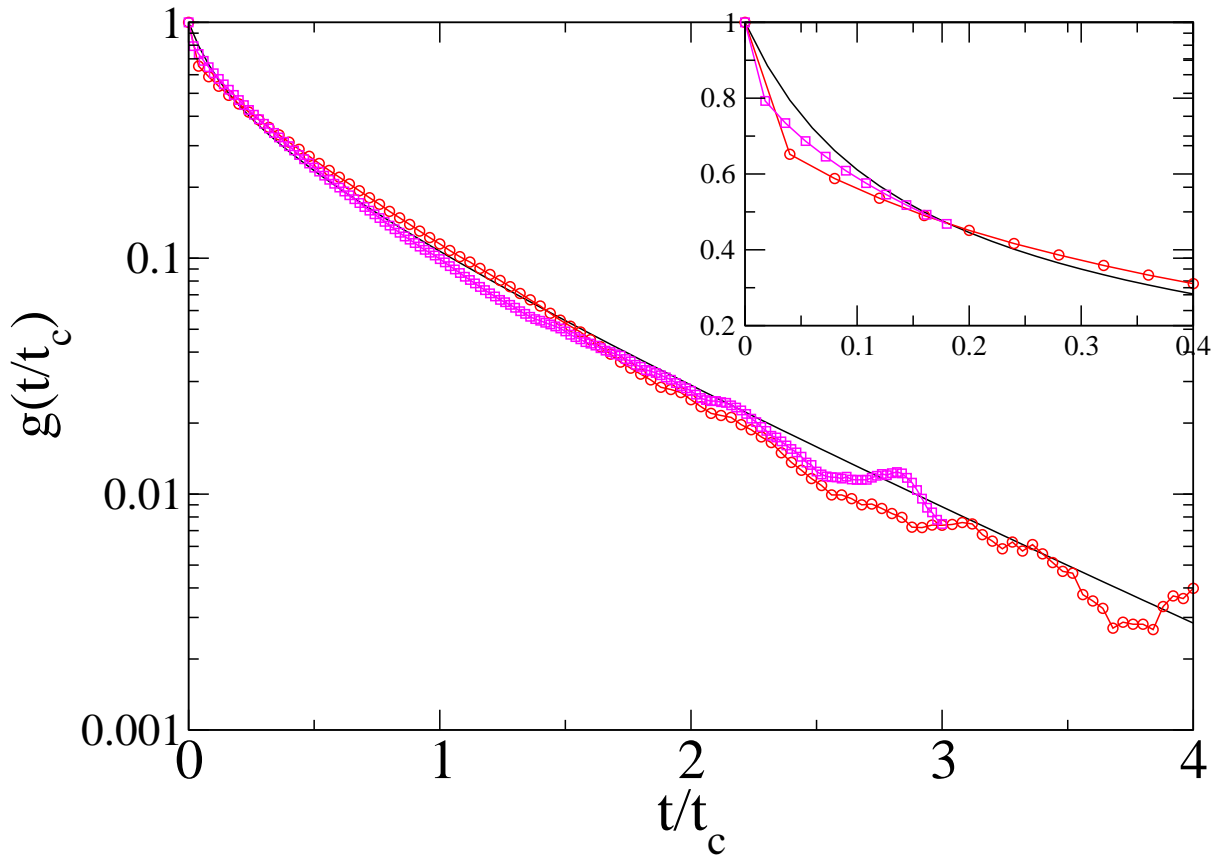}
       \caption{Height-height correlation functions, curve (\ref{a3}) and experimental 
points; circles refer to $\delta=0.04$ (II) and squares to $\delta=0.02$ (IV).
Insert: first 10 measured correlation points and curve (\ref{a3}).}
\label{height}
\end{figure}

Capillary wave theory for overdamped waves 
gives the following expression for the correlation function \cite{Volkert,Jeng}.
\begin{equation} \label{a3}
g(t/t_c) = \frac{2}{ \log (1+\kappa^2) } \int^\kappa_0 \, x dx 
\frac{\exp [-(x+x^{-1})(t/2 t_c)]}{1 + x^2}.
\end{equation}
The upper bound of the integral is $\kappa= 2 \pi  \xi / d$ with 
$\xi=\gamma/(g \Delta \rho)$ the correlation length and $d$ the diameter of the particles. 
The lower bound, determined by the size of the system, has been set equal to 0.
The upper bound is essential for the convergence of the integral and of influence
on the short time behavior of the correlation function.  
Cutting off the capillary waves at the short-wavelength side is the poor man's
way to handle the otherwise diverging interface width $\langle h^2 \rangle$. 
Fig. \ref{height} shows the function (\ref{a3}) together with data points referring to two experiments, 
denoted by II and IV \cite{Volkert}. The choice of $\delta$
is actually a fit of $t_c$. So we have plotted in the figure the data points
with $t/t_c=n \delta$ and $\delta=0.04$ for II and $\delta=0.02$ for IV.

Although the data points follow the curve reasonably well, the scatter of the 
data points is manifest and more important, though only visible in the insert, is 
the fact that the first few points are well below the curve. In this paper it is of less importance 
how well the experimental correlation function can be represented by a theoretical curve, 
since we are interested in the problem to directly calculate the persistence probabilities from the 
experimental correlation function.

\section{Persistence probabilities and sum rules}

To formulate the persistence probability for a discrete series, 
we form the $n \times n$ matrix $g_{i,j} = g_{|i-j|}$,
which is determined by the first $n$ values of $g_n$.
Using that the process is Gaussian, the probability on a sequence of values 
$(h_1, \cdots , h_n)$ is given by the formula
\begin{equation} \label{a4}
G (h_1, \cdots , h_n) =\frac{1}{ D^{1/2}}
\exp \left(- \frac{1}{2} \sum_{i,j} J_{i,j} h_i h_j \right), 
\end{equation}
with the normalization $D=(2 \pi )^n \det g$, where 
$\det g$ is the determinant of $g$. $J_{i,j}$ is 
the inverse of $g_{i,j}$, which is again a $n \times n$ matrix and  
its matrix elements also individually depend on $n$ (while those of $g$ 
only depend on $|i-j|$). The justification of (\ref{a4}) is based on the fact that the correlator 
$\langle h_i h_j \rangle$ as calculated with (\ref{a4}) leads indeed to $g_{|i-j|}$.

The expression for the persistence probability $p_n (h)$, on a sequence of
{\it precisely} $n$ events above $h$, follows as the ratio of two integrals 
\begin{equation} \label{a5}
p_n (h) = q^{-(n+)-} (h) /q^{-+} (h). 
\end{equation}
The numerator is the $(n+2)$-fold integral 
\begin{equation} \label{a6}
q^{-(n+)-} (h) = \int_{\cal D} \, dh_i \, G (h_0, \cdots , h_{n+1}),
\end{equation}
where the integration domain ${\cal D}$, indicated by the superscript on $q$, 
is given by the conditions $h > h_0 > -\infty,\,h < h_i < \infty, $ and $ h > h_{n+1} > -\infty$.
It selects the sequence of events: starting with a value below $h$, 
followed by $n$ points above $h$ and terminated by a value below $h$, 
which are the sequences of precisely $n$ successive values of the height $h_i > h$.
Similarly the denominator is given by the double integral
\begin{equation} \label{a7}
q^{-+} (h) =  \int^h_{-\infty} dh_0 \int^\infty_h \, d h_1 \, G (h_0, h_1)
\end{equation}
The inverse matrix $J_{i,j}$ depends for $n=2$ only on the first value $g_1$ 
of the correlation function. So the $q^{-+} (h) $ is generally given by 
\begin{equation} \label{a7b}
q^{-+} (h)  = \int^h_{-\infty} dh_0 \int^\infty_h \, d h_1 \, 
{\exp[-(h^2_0+h^2_1-g_1 h_0 h_1)/(1-g^2_1)] \over [2 \pi (1-g^2_1)]^{1/2}}
\end{equation} 
The denominator serves as normalization, since one has the relation
\begin{equation} \label{a7a}
q^{-+} (h) = \sum_{n=1} q^{-(n+)-} (h)
\end{equation} 
which gives the sum rule for the total probability 
\begin{equation} \label{a9}
\sum_{n=1} p_n (h)=1.
\end{equation} 
The derivation of (\ref{a7a}) is based on repeated expansion of the relation
\begin{equation} \label{a8}
q^{-+} (h) =  q^{-+-} (h) + q^{-++} (h) = q^{-+-} (h) + q^{-++- } (h)+ q^{-+++} (h) = \cdots ,
\end{equation} 
which simply states that the probability to find a sequence $-+$ is the same as
finding it from the events $-++$ and $-+-$. Together they extend the integration over
the last height variable over all values and the probability on $-+$ results. Repeatedly 
adding a new point in the sequence ending with $+$ leads to the identity (\ref{a7a}).

There exist one other sum rule. 
Consider $q^+ (h)$, which is the single integral over $h_0$ with $h_0>h$.
The expansion, similar to (\ref{a8}) starts as
\begin{equation} \label{a8a}
q^+ (h) = q^{+-} (h) + q^{++} (h) = q^{-+-} (h) + q^{++-} + q^{-++} (h) + q^{+++} (h).
\end{equation} 
Systematically replacing every $+$ at the begin or end of the string by the sum
of strings extended with a $+$ and $-$ gives
\begin{equation} \label{a10}
q^+ (h) = \sum_{n=1} \, n  \, q^{-(n+)-} (h).
\end{equation} 
This yields for the mean value the sum rule
\begin{equation} \label{a11}
\sum_{n=1}  n \, p_n (h)= q^+(h)/q^{-+} (h).
\end{equation} 
We have not found further sum rules. 

The sum rules (\ref{a9}) and (\ref{a11}), which generally apply for stationary Gaussian
processes, involve doable integrals. $q^{+} (h)$ is an
error function and $q^{-+} (h)$ can be reduced from  the double integral (\ref{a7b}) to a single
integral, for which an analytic expression exist for $h=0$ \cite{Hans}
\begin{equation} \label{a12}
q^{-+} (0) = {1 \over 2} - {1 \over \pi} \arctan \left({1+ g_1 \over 1 - g_1} \right)^{1/2}.
\end{equation} 
The sum rules are typical for the discrete series as the continuum limit for non-smooth
processes is singular. If one could straightforwardly define in the continuum limit a probability 
density $p(t)$ on a persistence interval $t$, one would expect that it would be related to the
discrete $p_n$ as
\begin{equation} \label{a13} 
p(n \Delta t) \Delta t \simeq p_n,
\end{equation} 
where we suppressed the $h$ dependence for the moment. The probability density
is then normalized
\begin{equation} \label{a14}
\int^\infty_0 dt \, p(t) \simeq \sum_{n=1} p_n =1.
\end{equation} 
Consequently one would expect the mean of the discrete series to diverge as
\begin{equation} \label{a15}
\sum_{n=1}  n \, p_n = (\Delta t)^{-1} \sum_{n=1}  (n \Delta t) \, p_n \simeq (\Delta t)^{-1} 
\int^\infty_0 dt\, t \,p(t) \sim (\Delta t)^{-1}. 
\end{equation} 
However the expression (\ref{a12}) shows, with $g_1 \simeq 1 - {\cal O} (\Delta t)$, that 
\begin{equation} \label{a16}
q^{-+} \sim (\Delta t)^{1/2},
\end{equation} 
while $q^+$ is independent of $\Delta t$. So the discrete mean diverges as $(\Delta t)^{-1/2}$
in contrast to the expected behavior (\ref{a15}). The  discrepancy is due to the Brownian 
fluctuations on all length scales for a non-smooth correlation function\cite{Krug,Volkert}. 
This gives a non-integrable probability density for short times, making the normalization
(\ref{a14}) questionable. 
 
We calculate the $p_n (h)$ and then check whether total and mean  corresponds with the 
exactly calculable ratios. In particular  (\ref{a11}) is a stringent test on the the calculated 
values of $p_n (h)$, since it more sensitive to the large $n$ values than ($\ref{a9})$.

The restricted integration domain prevents the integral (\ref{a6}) from 
straightforward evaluation. Although it involves 
a finite set of integrations, which indeed can be performed by standard 
techniques for small $n$, a direct evaluation of (\ref{a6}) is impossible for 
large $n$. As we shall show, expression (\ref{a6}) has a definite asymptotic 
large $n$ behavior, but the experimental data are not at all exclusively 
determined by this asymptotic behavior. In fact, the practical regime for 
which accurate data can be collected, shows important transient behavior.

\section{The Markovian case}

Our calculational scheme is inspired by the perturbation technique of Majumdar 
and Sire \cite{Sire}, which takes the Markovian case as lowest approximation. 
The Markovian case has an exponential decaying correlator 
\begin{equation} \label{b1}
g(t/t_c) = \exp (-\lambda t/t_c),
\end{equation} 
such that $g_n= g^n$ with $g=\exp(-\lambda \delta)$. The inverse matrix $J$ is
then a band matrix with all elements 0 except on the diagonal: 
\begin{equation} \label{b2}
J_{00}=J_{n+1,n+1}=1/(1-g^2), \quad \quad J_{ii}=(1+g^2)J_{00}, \quad \quad J_{i,i\pm 1}=-g J_{00}.
\end{equation} 
Note that this also holds for a  matrix of finite dimension $n$.
If the matrix $J$ is restricted to the diagonal and the subdiagonals, 
the corresponding joint probability can be factorized in several ways. We present here the
symmetric representation for the above Markovian case 
\begin{equation} \label{b3}
M (h_0, \cdots , h_{n+1}) = f_0 (h_0) \left( \prod^{n+1}_{i=1} K (h_{i-1}, h_i) \right) f_0 (h_{n+1}).
\end{equation}
The initial (and final) function is given by 
\begin{equation} \label{b4}
f_0 (x) = \exp [ -u_0 x^2] / [2 \pi]^{1/2} 
\end{equation}
with $u_0=1/4$. The kernel reads
\begin{equation} \label{b5}
K (x,y) = \frac{\exp [-u(x^2+y^2)+vxy] }{ [2 \pi (1-g^2)]^{1/2}},
\end{equation}
with the values 
\begin{equation} \label{b6}
u={1+g^2) \over 4(1-g^2)}, \quad \quad {\rm and} \quad \quad  v={g \over (1-g^2)}.
\end{equation}
Physically, the factorization results from the fact that in a Markovian process
the probability on an event only depends on the probability of the previous 
event. Mathematically, any matrix which is restricted to the diagonal and 
the subdiagonals, can be seen as a Markovian matrix. We use the more convenient symmetric form, 
allowed by time reversal symmetry, rather than the standard conditional probability,
with an asymmetric kernel. 
We will lean heavily on choosing the optimal Markovian approximation, which uses
optimal values for the $u$ and $v$ and not those tied in with $g$. 

Any of these representations give the values of $p_n (h)$ recursively.
Define a set of functions $f_n (x)$ with $f_0(x) $ given by (\ref{b4}).
$f_1 $ is constructed from $f_0$ as
\begin{equation} \label{b8}
f_1 (y) = \int^h_{-\infty} dx f_0 (x) K(x,y).
\end{equation} 
The general term is defined by recursion for $1 < i \leq n+1$ 
\begin{equation} \label{b9}
f_i (y) = \int^\infty_h dx \,f_{i-1} (x) \,  K(x,y).
\end{equation} 
Then $p_n (h)$ can be expressed as 
\begin{equation} \label{b10}
p_n  (h) = \int^h_{-\infty} dx f_{n+1} (x) f_0 (x) {\big/} q^{-+} (h).
\end{equation} 
In the Markovian case the multiple integral (\ref{a6}) becomes a repeated
integral transformation with $K$ as kernel. Asymptotically the result is 
dominated by the largest eigenvalue of the kernel, which then leads to a
persistence exponent. 

\section{Calculational scheme}

Our approximation scheme is based on a separation of $J$ in a Markovian 
part $M$ and a remainder $H$
\begin{equation} \label{c1}
J = {\cal M+H}. 
\end{equation}
$\cal M$ is a Markovian approximant, i.e. a matrix which is confined to the 
diagonal and the subdiagonals.  
Thus one can relate a kernel $K$ to $\cal M$  as in (\ref{b3}),  but with as yet
unspecified values of $u_0, u$ and $v$. Also a Markovian joint probability $M$ 
can be associated with this part (as in (\ref{b3}), which enables to define a 
Markovian approximation $p^0_n (h)$ 
\begin{equation} \label{c2}
p^0_n (h) ={1 \over q^{-+} (h) }  \int_{\cal D} \, dh_i M (h_0, \cdots, h_{n+1}), 
\end{equation} 
with $\cal D$ the same integration domain as in (\ref{a6}). The full probability $p_n (h)$ then reads
\begin{equation} \label{c3}
p_n (h) = p^0_n (h) \langle \exp (-{\cal H}) \rangle_0.
\end{equation} 
where the average $ \langle {\cal A} \rangle_0$ is defined as
\begin{equation} \label{c4}
\langle {\cal A}  \rangle_0  = { \int_{\cal D}  \, dh_i \, M (h_0, \cdots, h_{n+1} ) 
A (h_0, \cdots, h_{n+1})  \over \int_{\cal D} \,  dh_i \, M (h_0 , \cdots, h_{n+1}) }.
\end{equation} 
The exponential in (\ref{c3}) will be evaluated by the cumulant expansion. 
The Markovian average of the first cumulant of ${\cal H}$ as well as the 
higher cumulants are calculated in a similar iterative way as $p^0_n (h)$. 

There is a large freedom in choosing the Markovian part. 
Any matrix with only a non-zero diagonal and 
subdiagonals would do. This freedom can be exploited through the inequality
\begin{equation} \label{c5}
\langle \exp (-{\cal H}) \rangle \geq \exp (- \langle {\cal H} \rangle).
\end{equation}
The optimal Markovian matrix gives the largest value for the right-hand side.
The calculation thus consists of finding an optimal Markovian approximation and 
evaluating the corrections. For the convergence of the calculational scheme
the optimization is crucial.
As optimization parameters we can use $u_0$ and for each kernel a choice
for $u$ and $v$. The practical optimization is a trade-off between the
optimum and calculational simplicity. It is expedient to have all kernels
the same, with the exception of the first and last. They have anyway a 
different role, carrying the system from below $h$ to above $h$, while the other
progagators keep the system above $h$. So we remain with an initial (and final)
$u_0$, the first (and last) pair $u_1,v_1$ and the bulk pair $u,v$. A further
restriction stems from the chosen integration procedure. The integrals can
be fast and accurately evaluated using Gaussian quadrature, with a Gaussian
as weight factor. The initial  $f_0(x)$ is a Gaussian. 
Requiring that after an iteration the asymptotic behavior
is still the same Gaussian, $u,v$ and $u_0$ have to be related
as $u^2 = v^2 + u_0^2$. A constant Gaussian allows to use the same 
positions and weights in the Gaussian quadrature for each iteration. This
reduces the freedom from 5 to 3 parameters: $u_0,v_1$ and $v$.

Thus our calculational scheme starts out by finding for every $n$ the optimal values
of the parameters $u_0,v_1$ and $v$. According to the inequality (\ref{c5}) this gives
persistence probabilities $p_n (h) $ which are too low. Consequently the total
persistence and the mean as calculated with this approximation will be lower than 
the exact values calculated from the sum rules (\ref{a9}) and (\ref{a11}). Using the
the Gaussian quadrature this is a fast routine. The next step is the calculation of the
second cumulant, which leads to an overestimation of the persistence probability.
Thus including the second cumulant the sum rules are approached from above. 
As the second cumulant requires to calculate many correlation functions (going up
with $n^2$) this becomes already a longer calculation. It takes about an hour on 
a simple PC to calculate persistence probabilities up to $n=150$. 
Needless to say that the computation of the third cumulant, where the number of correlation
functions goes up as $n^3$,  is even more time consuming. In practice we could calculate the
third cumulant up to $n=40$ in a reasonable amount of time. Fortunately the changes
in the persistence probabilities due to the inclusion of the third cumulant are so small
that they do not influence the value of the sum rules. 

\section{Test of the approximation scheme}

Before we carry out the computation of the persistence probabilities of the actual data,
we inspect a clean case by taking the correlations points $g_n$ from the analytic 
expression (\ref{a3}). This has the advantage of noise-free 
input and allows to vary the sampling rate $\delta$. The test consists out of the
calculation of the sum rules with the cumulant expansion outlined in the previous section.
We consider two samplings of the curve (\ref{a3}), $\delta=0.04$ and twice as
narrow $\delta=0.02$, which correspond closely to the experiments II viz. IV \cite{Hans}). 
We take $\kappa=50$ in both cases. 
\begin{table}[h] \label{test}
\begin{center}
\begin{tabular}{|c|c|c|c|c|c|c|}
\hline
 & \multicolumn{3}{c|}{}& \multicolumn{3}{c|}{} \\*[-1mm]
 &\multicolumn{3}{c|}{II: \, $\delta=0.04$} &\multicolumn{3}{c|}{IV: \, $\delta=0.02$}\\*[2mm]
\cline{2-7}
 sum  & & & & & & \\*[-1mm]
   & 1st  & 2nd & exact & 1st  & 2nd & exact \\*[2mm]
  \cline{1-7}
   & & & & & & \\*[-1mm]
$\sum_n p_n (0)$ & 0.9935 & 1.00002  & 1.0000 
& 0.9932 & 1.0005 &  1.0000 \\*[4mm]
$\sum_n n p_n (0)$ & 4.6998 & 4.8517 & 4.8420  
& 6.3623 & 6.5417 & 6.5351 \\*[4mm]
\hline   
\end{tabular}
\caption{Sum rules for the testcurve; cumulants and exact value.} 
\end{center}
\end{table}
In Table 1 we have summarized the sum rules for the total (\ref{a9}) and 
and mean (\ref{a11}) persistence in the 1st and 2nd cumulant approximation, 
for the two samplings of (\ref{a3}) with level $h=0$ as discriminator. 
We have calculated 100 points for sampling II and 150 points for IV, 
which suffice to saturate the values of the sums. The third entry, exact,  gives the values
of the sum rules calculated from the ratios of integrals. 
The accuracy in the total probability according to the first cumulant, giving a rigorous 
lower bound, is amazing. For the mean value this is still
impressive, but indicates that the probability for the higher values of $n$ is 
somewhat too low. The second cumulant, always overestimating the probabilities,
makes well up for this deficiency of the lower bound. The overall impression is
that the scheme performs very well for points taken from the curve (\ref{a3}). 
We have doubled the cut-off $\kappa$ to see the influence. It marginally lowers 
the curve, except that $g_1$ is decreasing visibly. A larger 
cut-off makes the process less "smooth". This gives a deterioration of the 
bound. We have verified that including the third cumulant does not change 
the numbers in Table 1 appreciably.

\section{Comparison with the experiments}

The challenge is whether the calculation of the persistence probabilities from the data points 
of the correlation function is also accurate. First we test this again with inspecting
the sum rules, which are summarized in Table 2.
\begin{table}[h] %\label{exper}
\begin{center}
\begin{tabular}{|c|c|c|c|c|c|c|}
\hline
 & \multicolumn{3}{c|}{}& \multicolumn{3}{c|}{} \\*[-1mm]
 &\multicolumn{3}{c|}{state point II} &\multicolumn{3}{c|}{state point IV} \\*[2mm]
\cline{2-7}
 sum & & & & & & \\*[-1mm]
   & 1st  & 2nd & ``exact'' & 1st  & 2nd & ``exact'' \\*[2mm]
  \cline{1-7}
   & & & & & & \\*[-1mm]
$\sum_n p_n (0)$ &  0.9480 &   1.0024 & 1.0000 
&   0.9339  & 1.0034  &  1.0000 \\*[4mm]
$\sum_n n p_n (0)$ & 2.8760 &  3.7093 &   3.6513  
& 3.2761 &  4.9344 &  4.7933  \\*[4mm]  
\hline   
\end{tabular}  \label{exper}
\caption{Sum rules for the experiments; cumulants and calculated mean.}
\end{center}
\end{table}
\begin{figure}[h]
\begin{center}
   	\epsfxsize=\linewidth
	    \epsffile{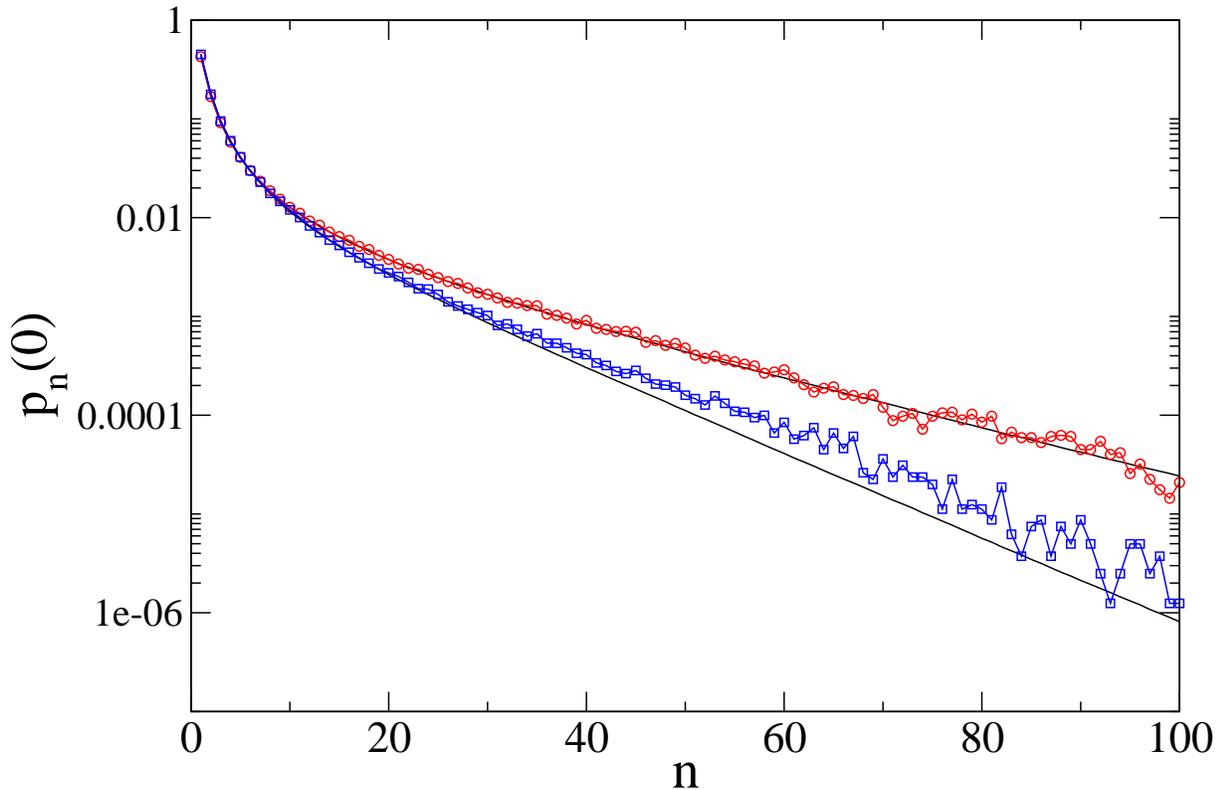}
\vspace*{4mm}

    \caption{Probabilities (lines) calculated from the measured correlation
function and measured probabilities 
(points) for state point II (lower curve, squares) and IV (upper curve, circles)
and $h=0$.}
\label{persis}
\end{center}
\end{figure}
The entry ``exact'' refers to the calculation of the sums using (\ref{a9}) and (\ref{a11}), 
which requires only the value of $g_1$. The marks around ``exact'' refer to the 
uncertainties in $g_1$ determining the ratios, which is not correlated to the uncertainty
in the other $g_n$, together determining the persistence probabilities. 
The sums for statepoint II are based 
on 100 points, which is about the number of reliable values of $g_n$. 
The sums of state point IV are extended to 150 points.
In both cases the value of $p_n (0)$ is then so small
that inclusion of further points will not change the sums. The lower bound
is considerably less accurate than for the generated $g_n$. This is not
surprising. The above mentioned scatter in the data adds to the non-Markovian
character. Even if the process were strictly Markovian, 
the noise would be seen as a deviation from Markovian behavior and would lead 
to a convergence question. The second reason is that the measured $g_1$ 
is quite low: $g_1=0.65$ (II) and $g_1=0.79$ (IV).
We noticed already that less ``smooth'' processes have a slower convergence.
However the second cumulant approaches nicely the correct values, indicating that
these calculated persistence probabilities are accurate. 

The real challenge is to see how well the persistence probabilities calculated from the measured
correlation points compare with measured probabilities. 
This is shown Fig. \ref{persis} for the two state points (with $h=0$).  
The calculated points are taken from the second cumulant approximation, since 
the third cumulant has no appreciable influence. The
agreement for state point IV is as good as one can hope for probabilities as small
as $10^{-5}$. For state point II there is a systematic deviation for larger $n$. 
This may be caused by artifacts resulting from the confocal slicing. 
$\langle h^2 \rangle$ and therefore the resolution relative to $\langle h^2 \rangle$ 
are higher for IV than for II, leading to more pronounced artifacts in II \cite{Volkert}.
Note that the asymptotic exponential decay sets in around $n=30$ which takes
as long as $15s$ in real time. 

\section{Discussion}

We have presented measurements for the correlation function and the persistence probability 
of the fluctuating heights of a colloidal interface. We have developed a 
calculational scheme which enables to find the persistence probability directly from
the measured correlation function. This means that our scheme is independent
of the agreement of the correlation function to a function like (\ref{a3}). 
The only assumption is that the distribution (\ref{a4}) is Gaussian, which 
stems from the fact that the thermal interface fluctuations are small 
deviations from equilibrium. The predictions of the persistence probabilities
from the measured correlation function agree very well with the measured
persistence probabilities.

Finally we like to make a few comments on the results.
\begin{enumerate}
\item Fig. \ref{persis} shows that the curve approaches an exponential decay, albeit that
a window from 30-100 points is not large for an accurate determination of the exponent. 
As we mentioned the two experiments may be seen as two samplings of the same 
correlation function. In order to compare the two samplings we translate the behavior to the 
time domain. For the two samplings we estimate the decay of $p_n (0) \sim \exp(- \psi(\delta)n)$.
Since $n \delta = t/t_c$, the persistence exponent $\theta$ equals 
$\theta=\psi(\delta)/\delta$. For the smallest $\delta$ we find $\theta=0.027$
and for the larger $\delta$, $\theta = 0.026 $. These two values need not be the same.
As observed by Ehrhardt et al. \cite{Ehrhardt},
the exponent for non-smooth processes is quite sensitive to the sampling rate.
The discrete series does not record what happens {\it in between} the measured points. 
The continuous process might dive below the level $h$ between recorded points. 
Such paths are excluded in a continuous formulation, but are included in the discrete version 
\cite{Majumdar2,Volkert}.  This leads to a smaller exponent for a larger $\delta$.
\item We note that we cannot take the continuum limit for practical and 
essential reasons. Making our $\delta$ smaller requires longer series to 
calculate as the two samplings show. But more 
importantly, a smaller $\delta$ will lead to a sharper peaked kernel, which 
ultimately has to approach a $\delta$-function. Such a delta peak cannot be treated 
by Gaussian quadrature, which has helped to speed up the integrations by about a factor 
thousand with respect to e.g. a Simpson rule. 
\item We have checked that higher cumulants have virtually
no influence on the persistence probabilities calculated here. 
We noticed, however, that although the  third cumulant is small, 
it develops a linear dependence on $n$. It has no influence 
on the presented sum-rule results nor on the presented persistence probabilities, 
since it changes only the probabilities which are already too small to contribute. 
But it will have a non-negligible influence on the decay exponent of the discrete series, 
showing up in a region which is beyond our measurements. This also means that the
experiments have not entered the fully developed asymptotic regime.
\end{enumerate}

{\bf Acknowledgement} The authors thank Wim van Saarloos and Henk 
Hilhorst for stimulating discussions.\\

\end{document}